\documentstyle[11pt]{aipproc}
\begin{document}
\pagestyle{empty}

\begin{tabular}{lr}
Visual Techniques Laboratory & \hspace{1.25truein} UWSEA PUB 94-05\\
Department of Physics, FM-15 &\\
University of Washington &\\
Seattle, WA 98195 USA &\\
\end{tabular}

\title{SUMMARY OF NUCLEAR AND PARTICLE ASTROPHYSICS SESSIONS
\thanks{To be published in {\it Proc. 5th Conference on Intersections
of Particle and Nuclear Physics.}} }

\author{ R. Jeffrey Wilkes \thanks{Supported by DOE Grant DE-FG06-91ER40614.}\\
Department of Physics, FM-15\\
University of Washington\\
Seattle, WA 98195 USA\\
}
\date{}

\maketitle

\begin{abstract}
Astrophysics is gaining increased attention from the particle and
nuclear physics communities, as budget cuts, delays, and
cancellations limit opportunities for breakthrough research at
accelerator laboratories.  Observations of cosmic rays (protons
and nuclei), gamma rays and neutrinos present a variety of
puzzles whose eventual solution will shed light on many issues
ranging from the nature of fundamental interactions at extreme
energies to the mechanisms of astrophysical sources.  Several
important detectors are just beginning full-scale operation and
others are beginning construction.
\end{abstract}

\section*{INTRODUCTION}
Astrophysics has become a growth industry in the past few years,
fuelled partially by new and exciting results from projects like
the Gamma Ray Observatory (GRO), partially by the diminished
prospects for future accelerator-based research, and partially by
the expectation that new astrophysics projects about to come on
line or beginning construction (DUMAND, SuperKamiokande, SNO,
MILAGRO) will provide at least partial answers to some long-standing
questions.

We heard reports on a wide variety of activities, ranging from studies of
cosmic rays (protons and nuclei), to the full range of the gamma ray
spectrum (ie, from MeV to EeV), to neutrino detectors of all species. This
brief summary is intended as a guide to the sessions; see papers in this
volume by the persons named, for details, illustrations, and references.

\section*{GAMMA RAYS}
The BATSE experiment aboard GRO has provided our first really
good systematic look at gamma ray bursts: transient emissions in
the Mev--GeV range lasting from msec to minutes, with no
simultaneous emissions below the X-ray band, and no quiescent
emission from the same source at any energy. David Band reported
the latest results, which indicate that while the sources are
directionally isotropic within statistics, their cumulative
intensity distribution deviates from the $-3/2$ power law
expected for homogeneity in 3-space. This result favors the
hypothesis that the bursters are distributed at cosmological
rather than Galactic distance scales.

In a higher energy regime, the CYGNUS experiment at LANL has been
studying potential gamma ray sources in the TeV--PeV range using
relatively conventional extensive air shower (EAS) techniques,
supplemented by some very cost effective water Cerenkov detectors
made from commercial backyard swimming pools. As its name
implies, the original motivation for CYGNUS was to investigate
the muon-rich (hence presumably hadronic) EAS reportedly
emanating from the direction of Cyg X3 in synch with its radio-
frequency periodicity. Unfortunately, shortly after the first
round of inadequately prepared experiments got on line, Cyg X3
shut down, perhaps due to budget cuts, and has not been heard
from since. This is not surprising, since we don't fully
understand all the long term periodicities to which our own Sun
is subject, and it may be that we will hear from Cyg X3 (and Her
X1 and all the other hopefuls) again. Meanwhile, the CYGNUS
detector's area has been built up to 80,000 m$^2$ through careful
incrementalism, with threshold 70 TeV and 0.5$^o$ resolution, and
the result still appears to be ``no significant observations of
TeV gamma ray point sources". Similar negative results are
reported from searches for AGNs, all-sky surveys, periodicity
scans on known lower-energy sources, and searches for Primordial
Black Holes (PBHs).

Jim Matthews reported on CASA-MIA, the giant (480 meters square),
beautifully instrumented air shower array  centered on the Fly's
Eye II site in Utah. They have proven the $<1^o$ resolution of
their detector by observing gamma-ray shadows cast by the sun and
moon. Searches for 100 TeV gamma emission from point source
candidates like Cyg X3, Her X1 and Crab are all negative, as are
searches for signals from EGRET sources like Mk 421, and for 30
BATSE sources. With full area in operation, they expect to be
able to test predictions of diffuse Galactic gamma ray
intensities after a few years running.

Let me indulge in a brief editorial: the continuing saga of
negative results from point source searches should be {\it
en}couraging, not {\it dis}couraging, for we know that the Galaxy
is swarming with cosmic rays above the TeV range, and hadronic
interactions, in the sources and in the interstellar medium, have
to produce pions that decay to TeV gamma rays (and neutrinos) --
{\it unless} something funny happens to the hadronic interaction
at ultra-high energies, which is becoming very hard to believe
given current accelerator results, but would be great fun.  So
either we are merely having exceptionally bad luck regarding the
correlation of source periodicities and human scientific history,
or we are about to learn something fundamental about source
mechanisms or even the hadronic interaction. We should remain
cheerful.

Gus Sinnis told us about plans for MILAGRO, a large-area water
Cerenkov detector under construction near LANL. The existing pond
at 8600' altitude has surface area 5000$m^2$, and will be
instrumented with 750 PMTs and 200 scintillation counters. The
resulting detector will have a threshold of only 250 GeV. The
project will begin data taking in 1997/8 and will provide the
first TeV range all-sky survey, as well as excellent sensitivity
for observations of GRBs and AGNs, and solar physics studies.

Cy Hoffman, in a separate paper, pointed out that MILAGRO will be
able to observe a significant signal from evaporating PBHs, with
sensitivity about 400 times that of CYGNUS.

\section*{COSMIC RAYS}
Tom Gaisser outlined the theoretical issues for cosmic ray
studies above 100 TeV. Our goal is to elucidate processes
producing the ``knee" (steepening of the spectral slope at $\sim
10^{15}$ eV) as well as the exciting prospect of exploring the
``ankle" (increased slope) region at $10^{17-20}$ eV, where
extragalactic sources come into play. To do this we need to
properly extrapolate accelerator results from $\surd{s} \sim 1$
TeV and the central rapidity region into these enormous energies
and the extreme forward direction (which is all that ground-based
detectors will observe). Recent models suggest that the
composition (ratio of various nuclei to protons) is energy
dependent for a single supernova, but depends also upon the
mix of stellar types present in the Galaxy. Extensive long-term
studies of the primary spectrum and composition will be needed.

Meanwhile, as Yoshi Takahashi reported, JACEE (Japanese-American
Cosmic ray Emulsion Experiment) is chipping away at the problem
with incremental balloon flight exposures of emulsion chambers,
building up an impressive database which is gradually approaching
the knee from below. This experiment has the great advantage of using a
vertex detector, so primary particles are directly observed and
identified. At TeV/nucleon energies, spectra of
heavy nuclei (Z$\geq 6$) are flatter than the proton spectrum, while
the He/p ratio is twice that observed at 100 GeV, suggesting a
heavier composition near the knee, as predicted by some models.
Improvements in balloon flight capabilities make possible an explosive
growth in the rate of data acquisition, given adequate support.

Gene Loh told us about observations at the other end of the
graph, from Fly's Eye, the atmospheric scintillation detector in
Utah which looks at the very highest energy cosmic rays due to
its enormous effective area. They recently recorded an event at
$3 \times 10^{20}$ eV (about 50 J!). Their results on the
spectrum and composition are consistent with a 2-component theory
in which the Galactic flux (below $3 \times 10^{18}$ eV) has a
heavy (iron-rich) composition while a light (proton-rich)
composition applies to the extragalactic component above that
energy.

\section*{NEUTRINO EXPERIMENTS (RUNNING)}

Ken Lande and collaborators are continuing to operate the
Homestake chlorine detector system established by Ray Davis
nearly 25 years ago. Paul Wildenhain described in detail the
procedures used to ensure consistent results over this very long
period of operation. There is no indication of a significant
change in the multi-year average flux of solar neutrinos over the
lifetime of the experiment.

Jeff Nico presented the latest update from the Russian-American
Gallium Experiment, which retains the acronym SAGE despite the
departure of ``Soviet" from the political vocabulary. Analysis of
SAGE-I (30 tons of Ga) is complete and SAGE-II (55 tons) is still
running, but equipment upgrades and careful procedures to
minimize effects of backgrounds still produce a result which is
0.56--0.60 of the SSM prediction.

Results from the MACRO experiment at Gran Sasso were presented by
Ed Kearns. The MACRO flux limit for monopoles is approaching the
Parker limit. Observations of upward-going muons allow MACRO to
check the atmospheric neutrino $(\nu_\mu/\nu_e)$ ratio.
Kamiokande and IMB reported apparent deficits relative to model
predictions which might indicate neutrino oscillations. MACRO
observes a deficit which does not have sufficient statistical
significance to exclude the no-oscillations hypothesis.

Results from the LSND accelerator experiment at Los Alamos on
possible neutrino oscillations were apparently too preliminary to
go on the record, but this experiment may prove worth watching.

\section*{NEUTRINO EXPERIMENTS (UNDER CONSTRUCTION)}

Ken Young described the progress made on DUMAND, which has finally become
a reality after so many years of planning and detector development. In
December, 1993, the basic infrastructure for the DUMAND experiment was put
in place on the ocean bottom off the Big Island of Hawaii: the underwater
junction box, shore cable, environmental monitoring equipment, and the
first string of phototubes. Although the string controller electronics
module developed a leak after 10 hours of operation, sufficient data were
collected to prove successful operation of the experiment and even
reconstruct some muon tracks. The problem has been fixed and three strings
are to be deployed later this year or early next year, as soon as suitable
ship support can be arranged.

Two major new solar neutrino experiments are in the early stages
of construction, both with a 1996 turn-on date. Bhaskar Sur
described progress on SNO, with emphasis on the delicate problems
of background reduction and calibration. At 6800 mwe depth, and
instrumented with 9500 PMTs, the SNO detector will have 1000 T of
heavy water inside a vessel containing 7200 T of light water.
This will provide the unique capability of observing neutral
current interactions.

Todd Haines described the capabilities of Super-Kamiokande, which
will come on line at about the same time as SNO. With a 22,000 T
fiducial volume and over 13,000 PMTs, this huge detector should
record over 8000 events per year with a 5 MeV threshold, and push
proton decay lifetime limits into the $10^{34}$ yr range. A
supernova at 10 kpc would produce over 5000 events.

\section*{PROSPECTS}

The next Intersections Conference is guaranteed to see some exciting
results from the many existing projects now reaching full capability as
well as the new experiments just getting under way. Several speakers
explicitly promised solutions of long-standing problems by the next
conference, so we shall hold them to it! Thanks are due to my
co-coordinator Gina Rameika, and all of the speakers mentioned above, for
making these sessions a success.

\end{document}